# Energy-Based Sequence GANs for Recommendation and Their Connection to Imitation Learning

**Jaeyoon Yoo**
ECE, Seoul National University
yjy765@snu.ac.kr

**Heonseok Ha**
ECE, Seoul National University
heonseok.ha@gmail.com

**Jihun Yi**
ECE, Seoul National University
t080205@snu.ac.kr

**Jongha Ryu**
ECE, UC San Diego
jongha.ryu@gmail.com

**Chanju Kim**
NAVER Corp.
chanju.kim@navercorp.com

**Jung-Woo Ha**
NAVER Corp.
jungwoo.ha@navercorp.com

**Young-Han Kim**
ECE, UC San Diego
yhk@ucsd.edu

**Sungroh Yoon**
ECE, Seoul National University
sryoon@snu.ac.kr

## ABSTRACT
Recommender systems aim to find an accurate and efficient mapping from historic data of user-preferred items to a new item that is to be liked by a user. Towards this goal, energy-based sequence generative adversarial nets (EB-SeqGANs) are adopted for recommendation by learning a generative model for the time series of user-preferred items. By recasting the energy function as the feature function, the proposed EB-SeqGANs is interpreted as an instance of maximum-entropy imitation learning.

## 1 INTRODUCTION

Over the past decades, numerous attempts have been made to build efficient mechanisms to recommend items (such as products, songs, videos, news articles, and so on) to users based on the preference of individual users, with the ultimate goal of increasing the level of satisfaction for users and the corresponding level of revenue for providers. Due to their high computational cost, traditional recommendation systems (see, for example, [17]) have typically dealt with relatively small datasets. Availability of large-scale datasets as well as advances in hardware for computation and storage, however, are changing the landscape for development of more powerful recommendation systems that have both high accuracy and scalability [5, 11].

In this context, recommendation algorithms based on deep learning have recently received much attention. A recommendation system can be viewed as a *mapping* from a sequence of items a user preferred to a new item the user is likely to prefer. As a universal approximator[15], a deep neural network is expected to learn this mapping efficiently and accurately in a scalable manner. Moreover, deep learning often finds proper representations of items implicitly, thus allowing transformative recommendation system designs across multiple domains. As a consequence, several variations of deep neural networks, including autoencoders [27], convolutional neural networks(CNNs) [18], and recurrent neural networks(RNNs) [11], have been proposed for recommendation. In order to emulate typical behaviors of a user more explicitly, a generative model can be utilized to reproduce a sequence of items preferred by the user. For example, a hidden Markov model has been trained to generate a next item from previous items and recommend it to the user [26].

In this paper, we proposes to use state-of-the-art generative adversarial networks (GANs) [9] as a deep learning based generative model for recommendation. More concretely, we combine energy-based GANs(EBGANs) [35] and sequence GANs(SeqGANs) [34] to develop energy-based sequence GANs (EB-SeqGANs) to learn the sequential preference of a user and use the resulting generative deep network to produce a recommended next item to the user. The use of GANs in general and EB-SeqGANs in particular (which, to the best of our knowledge, has not been explored before) provides a promising framework for the task of recommendation for the following two reasons.

First, GANs harness both the descriptive and discriminative power of deep learning effectively to train a generative model with high accuracy and scalability. Since the seminar paper by Goodfellow et al. [9], GANs have proved to outperform existing algorithms in many tasks ranging from face image generation [25] to robust control [13]. GANs, properly trained, are expected to predict a precise pattern of each user's preference.

Second, EB-SeqGANs, with a proper choice of the potential function, can be interpreted as an instance of imitation learning based on the maximum entropy principle. Imitation learning aims to find a suitable decision by mimicking the experts' demonstration. Consequently, the proposed EB-SeqGANs are expected to be well trained from expert recommendations (in this case, the actual items preferred by the users themselves).

The rest of the paper is organized as follows. Section 2 provides a brief introduction on GANs and imitation learning. In Section 3, we develop our recommendation system using EB-SeqGANs. Section 4 is devoted to elucidate the connection between EB-SeqGANs and imitation learning. Section 5 introduces related work. In section 6, we discuss the ideas presented and conclude.

Throughout the paper, music recommendation will be used to illustrate in concrete terms how EB-SeqGANs can be used for recommendation. The proposed approach, however, can be applied more broadly and easily transformed to other domains.

## 2 BACKGROUND

### 2.1 Generative Adversarial Networks

Generative adversarial networks [9] is a class of generative models that learns by game theoretic competition between a generator $G$ and a discriminator $D$. The generator $G$ learns the underlying distribution $p_{\text{data}}$ based on data, so that $G$ can generate an artificial data that resembles the real data. On the other hand, the discriminator $D$ learns to distinguish between the true data and the fake one generated by generator. $G$ and $D$ are usually implemented with deep learning models. Once the learning procedure is done, one would expect that $G$ may have enough generative power to fool discriminator, which implies that $G$ generates a data looks like the real one. The following is the objective for GANs:

$$\max_{D} \quad \mathbb{E}_{S \sim p_{\text{data}}} \log D(S) + \mathbb{E}_{S \sim p_G} \log(1 - D(S)) \quad (1)$$

$$\min_{p_G} \quad \mathbb{E}_{S \sim p_G} \log(1 - D(S)) \quad (2)$$

$S$ is data on the domain of interest. It may be a two dimensional matrix in the image domain and a sequence of items in recommender system domain. $D(S)$ represents the likelihood for discriminator $D$ to predict $S$ to be real data. While $p_G$ is the distribution from which generator $G$ generates a fake data, $p_{\text{data}}$ is the distribution of the real data. $\mathbb{E}_{S \sim q}[f(S)]$ represents an expectation of $f(S)$ when $S$ follows the distribution $q$. (1) means discriminator $D$ must increase the likelihood of real data and decrease the likelihood of the one that generated by generator. At the end, this leads $D$ to distinguish the real data from the fake one. (2) means that the generator $G$ should be optimized to generate the data that looks real so that the discriminator $D$ gives high likelihood to the generated data. Then optimizing generator $G$ leads $p_G$ to be optimized to $p_{\text{data}}$ and vice versa. Thus we write $p_G$ as an optimization variable in (2). In the rest of the paper, a generator refers to either $G$ or $p_G$ with abuse of terminology. We define $S$ as a data but we would like to remark that $S$ can be used to mean a state in the section 2.4 and 4.

### 2.2 Energy-Based Generative Adversarial Networks

Energy-based GANs [35] is one of many variants of GANs. It changes the meaning of discriminator's output $D$ from the likelihood in (1), (2) to *energy* function, and $D$ learns to make the energy function low for data from $p_{\text{data}}$ and high for those from $p_G$. More precisely, the objective is given as follows:

$$\min_{D} \quad \mathbb{E}_{S \sim p_{\text{data}}} D(S) + \mathbb{E}_{S \sim p_G}[m - D(S)]^+ \quad (3)$$

$$\min_{p_G} \quad \mathbb{E}_{S \sim p_G} D(S) + \psi(p_G) \quad (4)$$

$D(S)$ means the energy for data $S$ rather than likelihood as in previous section. This difference attributes to the change from (1),(2) to (3),(4). Discriminator discriminates real data from fake data by minimizing the energy of data from $p_{\text{data}}$ and maximizing that of data from $p_G$. This is described in (3). Generator is optimized to generate samples with low energy, in other words, realistic data in (4). $[x]^+ = \max\{x, 0\}$ denotes a margin function, and this term in updating $D$ cuts off the fake data from $G$ when it is too unrealistic to $D$, i.e. has energy larger than $m$. Also $\psi_G(\cdot)$ is a functional of distribution and plays a role of preventing mode collapsing of $p_G$. Without this term, $p_G$ would collapse down to only one datum that minimizes $D$.

### 2.3 Sequence generative adversarial networks

Sequence GANs is proposed to deal with sequential data in GANs framework [34]. It still uses the same objective in (1) for $D$, but considers a temporal structure of data when optimizing $p_G$. We use the following gradient of the objective for $p_G$ when the data is of the form $S = s_1 s_2 \ldots s_T$:

$$\sum_{t=1}^{T} \mathbb{E}_{s'_1 s'_2 \ldots s'_t \sim p_G(S)} \nabla p_G(s'_t | s'_1 \ldots s'_{t-1}) Q(s'_1 \ldots s'_t). \quad (5)$$

In (5), $Q(s_1 \ldots s_t)$ denotes the expected log-likelihood of the discriminator given $s_1 \ldots s_t$. The detail to derive (5) is close to (9)-(14) in the next section, so we omit the derivation of (5). For the interested reader, please refer to [34]. While the learning procedure, we sample $s_1 \ldots s_t$ and then estimate $Q(s_1 \ldots s_t)$ from rollout samples. Then, we optimize $p_G$ using the estimated $Q$ from (5). Updating $D$ is to optimize (1).

### 2.4 Maximum Entropy-based Imitation Learning

Imitation learning is an algorithm to solve decision making process [1]. The set of disciplined actions at each state is called policy. The basic idea of imitation learning is to solve the problem by mimicking the expert's policy based on given demonstrations.

Ziebart et al. introduced Maximum entropy(MaxEnt) imitation learning which is based on the maximum entropy principle [16] [36]. In [36], Ziebart et al. regarded an imitation learning as finding a policy $P$ which satisfies

$$\mathbb{E}_P[f(S, A)] = \mathbb{E}_{P_E}[f(S, A)], \quad (6)$$

where $f(S, A)$ is a feature vector for a state $S$ and an action $A$, and $P_E$ denotes the expert's policy which is approximated by the given demonstration. However, this is still an ill-posed problem in a sense that we cannot pick one policy among many candidates satisfying above. To give the guideline of the choice, [36] adopts the maximum entropy principle which states that the most reasonable probability distribution with constraints is the one that maximizes its entropy [16]. More precisely, the problem now can be formulated as

$$\min_{P} -H(P)$$
$$\text{s.t. } \mathbb{E}_P[f(S,A)] = \mathbb{E}_{P_E}[f(S,A)] \tag{7}$$

where $H(P)$ denotes the entropy of $P$.

Due to Slater's condition strong duality holds, and one can obtain primal and dual solutions from KKT condition [4]. The solution has the form

$$P(S,A) = \frac{\exp(-c^T f(S,A))}{Z}, \tag{8}$$

where $c$ is a vector that maximizes the log-likelihood of expert's demonstration, and $Z$ is a partition function. There is freedom to choose a feature vector $f(S,A)$, and domain knowledge may be exploited depending on applications. For example, for a sequential recommender system, one possible option is using a feature from word2vec [22] learning inspired by natural language processing [21]. In this paper, for simplicity we choose a state-action occurrence as our feature. A state-action occurrence feature has the same dimension with the number of state-action pairs. Provided the order of the state-action pairs, the $i$th element of feature vector, $f_i(S,A)$ is 1 when $(S,A)$ is a $i$th state-action pair and 0 otherwise. In this case the solution can be represented as $P(S,A) = \exp(-c(S,A))/Z$.

## 3 ENERGY-BASED SEQGANS FOR RECOMMENDATION

In this section, we would propose generative adversarial networks adopted recommender system specified in Algorithm. 1. We apply EBGANs [35] to recommender system which gives the form of line 6-11 in Algorithm. 1. EBGANs is known to be robust to hyperparameters and has more stable training process [2] compared to the original GANs in (1) and (2). Considering the temporal structure, we extend EBGANs to EB-SeqGANs for recommender systems. This extension adds line 1-4 in Algorithm. 1.

To apply EBGANs to a recommender system, we need to specify the data and the models for the generator and the discriminator. First, in music recommendation system, the data $S$ in (3) and (4) may correspond to a played history. We write $S = s_1 s_2 s_3 \ldots s_T$ where $s_t$ represents the song played at timestep $t$. Secondly, for the generator $G$, as a music recommender system exploiting a listening history must be causal, we use recurrent neural networks (RNNs), which has showed an outstanding performance for temporal data processing [21] [29]. For the discriminator $D$, any deep learning model such as 1D convolutional neural networks or autoencoder can be used [35].

We use a negative entropy $-H(p_G)$ for $\psi$ in (4) instead of the heuristic repelling regularization term in [35]. As the negative entropy becomes smaller, $p_G$ is less likely to be sharp, and mode collapsing is avoided. The validity of using the negative entropy will be corroborated by a theoretical interpretation in the next section.

Considering the temporal structure of $p_G$, we can convert the update process of $p_G$ in (4) into SeqGANs framework [34]. Substituting negative entropy in (4), the gradient of the objective of $p_G$ is given as follows.

**Algorithm 1** EB-SeqGANs($p_G$, $D$)

1: **procedure** SeqGANs($p_G$, $D$)
2:     $g \leftarrow \sum_t \sum_{s_1 \ldots s_t} p_G(s_1 \ldots s_t) \nabla \log p_G(s_t | s_1 \ldots s_{t-1})$
3:     $\sum_{s_{t+1} \ldots s_T} p_G(s_{t+1} \ldots s_T | s_1 \ldots s_t)[D(S) + \log p_G(S)]$
4:     optimize $p_G$ with $g$ as the gradient
5:
6: **Require** : 1) Choose $m$, the margin value. 2) Choose the regularizer term $\psi$(negative entropy in this paper), 3) Obtain data sampled from $p_{\text{data}}$
7: **Initialize** $p_G$ and $D$
8: **repeat**
9:     update $D$
        maximize $\mathbb{E}_{S \sim p_{\text{data}}} D(S) + \mathbb{E}_{S \sim p_G}[m - D(S)]^+$ w.r.t $D$
10:    update $p_G$
        minimize $\mathbb{E}_{S \sim p_G} D(S) + \psi(p_G)$ w.r.t. $p_G$
        by SeqGANs($p_G$, $D$)
11: **until** convergence

$$\nabla \mathbb{E}_{p_G(S)}[D(S) + \log p_G(S)] \tag{9}$$
$$= \sum \nabla p_G(S)[D(S) + \log p_G(S)] + \sum p_G(S) \nabla \log p_G(S) \tag{10}$$
$$= \sum \nabla p_G(S)[D(S) + \log p_G(S)] + \sum \nabla p_G(S) \tag{11}$$
$$= \sum p_G(S) \nabla \log p_G(S)[D(S) + \log p_G(S)] \tag{12}$$
$$= \sum p_G(S) \sum_t \nabla \log p_G(s_t | s_1 \ldots s_{t-1})[D(S) + \log p_G(S)] \tag{13}$$
$$= \sum_t \sum_{s_1 \ldots s_t} p_G(s_1 \ldots s_t) \nabla \log p_G(s_t | s_1 \ldots s_{t-1})$$
$$\sum_{s_{t+1} \ldots s_T} p_G(s_{t+1} \ldots s_T | s_1 \ldots s_t)[D(S) + \log p_G(S)] \tag{14}$$

Note that While obtaining (12), we change the order of summation and gradient and use the fact that the sum of probability is 1. In (14), $\sum_{s_{t+1} \ldots s_T} p_G(s_{t+1} \ldots s_T | s_1 \ldots s_t)[D(S) + \log p_G(S)]$ is expected value of $D(S) + \log p_G(S)$ given $s_1 \ldots s_t$ and plays the same role with $Q$ in (5). Thus, for the $p_G$ with temporal structure, (4) can be optimized as in SeqGANs. Consequently, each generator and discriminator optimization would follow (14) and (3), respectively. We would like to call this update process by energy-based sequence generative adversarial networks(EB-SeqGANs).

Once EB-SeqGANs in a recommender system is trained, we can exploit the learned model to recommend items by providing users a sequence that $p_G$ generates. The overall process contains two phase as illustrated in Figure 1. In the training phase, we train generator and discriminator with EB-SeqGANs framework given users' histories. In the recommendation phase, the learned generator is exploited to generate the recommending list given the songs that user listened before.

## 4 CONNECTION BETWEEN EBGAN AND IMITATION LEARNING

In this section, we show that how to perform imitation learning in recommender system, and as a result the energy-based GANs can be interpreted as imitation learning. The overall flow of this section is illustrated in Figure 2.

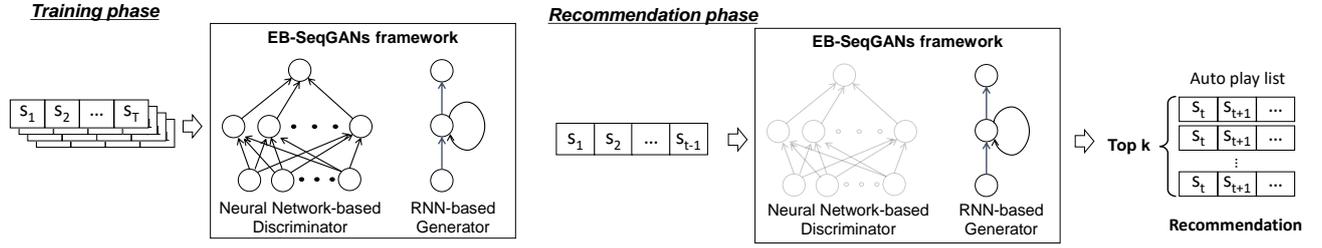

Figure 1: Overall process of EB-SeqGANs in recommender system

Music recommendation can be viewed as a series of decision making process in that it determines which item to suggest. That is, it is to determine which song to recommend at timestep $t$, provided that the user listened $s_1, s_2, s_3, \ldots, s_{t-1}$ before.

The common framework for the decision making process is reinforcement learning [30]. Reinforcement learning finds a policy that maximizes the reward the agent gets by performing an action $a$ at each state $s$. There have been several papers that proposed an application of reinforcement learning in recommendation system [8] [31]. Music recommendation problem can be formulated as a reinforcement learning problem by considering a history of played songs as a state and recommending a song to listen as an action. Assuming the recommended song must be played at each time, the transition is deterministic. In addition, a reward would be a numeric value that indicates how much the user likes the recommended song $a$.

A reward-maximizing policy found by reinforcement learning can be regarded as a music recommender system. However, defining the reward practically is hardly possible. Thus it is not straightforward to apply reinforcement learning algorithms to music recommender system.

Instead, recommender system stores the users' histories. Thus, imitation learning is a reasonable approach to construct the recommender system.

As mentioned above, the solution of MaxEnt imitation learning is the form of $P(S,A) = \exp(-c(S,A))/Z$, and $c$ is found by maximizing the log–likelihood of the experts' demonstration. Several attempts show how to implement the optimization of $c$ [36] [24] [33] [7]. While most of them are not scalable, the method in [7] shows its scalability. However, since [7] used the gradient which has a fraction form, it causes overflow and high variance issues in that denominator part of the gradient is a probability density lower than 1. However, as we will show, our method described in (20), (21) does not have such form, thus may not suffer from those issues.

To optimize $c$ to maximize the log–likelihood of the experts' demonstration, we will use gradient-based optimization and the gradient of the log–likelihood is given as follows.

$$\nabla \log(\text{likelihood}) = -\mathbb{E}_{(S,A)\sim\text{demo}}\nabla c(S,A) + \mathbb{E}_{(S,A)\sim P(S,A)}\nabla c(S,A) \quad (15)$$

In the first term, *demo* stands for the expert's demonstrations. The first term can be calculated easily by substituting given demonstrations. Second term is, however, intractable for the system with large state and action space as well as long timestep. It is equivalent to the negative term in Restricted Boltzmann Machine (RBM) [12], and hence this term can be estimated by the sampling method used in Restricted Boltzmann Machine. One can use importance sampling, but it raises high variance. Using metropolis-hasting (MH) or Gibbs sampling may be another option, however, these methods have high computational complexity as they require a large number of inferences.

To circumvent those disadvantages, we introduce an optimization variable $q(S,A)$ in place of the sampling distribution $P(S,A)$ in (15). $q(S,A)$ can be of any family of distribution from which it is easy to sample. Using $q(S,A)$ and approximating the gradient of log–likelihood brings out the following equation.

$$\nabla \log(\text{likelihood}) \approx -\mathbb{E}_{(S,A)\sim\text{demo}}\nabla c(S,A) + \mathbb{E}_{q(S,A)}\nabla c(S,A) \quad (16)$$

Hence, by integrating the above equation, we can find an approximate log–likelihood(LL) up to constants.

$$\text{Approximate LL} = -\mathbb{E}_{(S,A)\sim\text{demo}}c(S,A) + \mathbb{E}_{(S,A)\sim q(S,A)}c(S,A) \quad (17)$$

As we introduce $q(S,A)$ instead of $P(S,A)$ to ease the sampling procedure, we need to impose the constraint that $q(S,A)$ is close to $P(S,A)$. Hence we require that $q(S,A)$ and $P(S,A)$ are close in KL divergence. It can be seen as approximate inference [3]. Still, however, there is freedom to choose among $KL(q||P)$ and $KL(P||q)$ since KL divergence is asymmetric. We choose $KL(q||P)$ to be minimized for the following reason. First, $KL(P||q)$ requires sampling from $P(S,A)$, which is desired to be avoided. Second, when $P(S,A)$ is multimodal, which is a common case, $KL(P||q)$ would average out the modes and result in a poor approximation [3].

Summing up, we can modify the original problem that optimizes $c$ with the gradient (15) by the following two-step problem.

$$\min_{q} \text{KL}(q||P) \text{ for fixed } c \quad (18)$$

$$\max_{c} -\mathbb{E}_{(S,A)\sim\text{demo}}c(S,A) + \mathbb{E}_{(S,A)\sim q(S,A)}c(S,A) \text{ for fixed } q \quad (19)$$

Ideally, if $q$ minimizes KL divergence, $q(S,A)$ matches up $P(S,A)$ exactly and the gradient of approximate LL becomes the exact one.

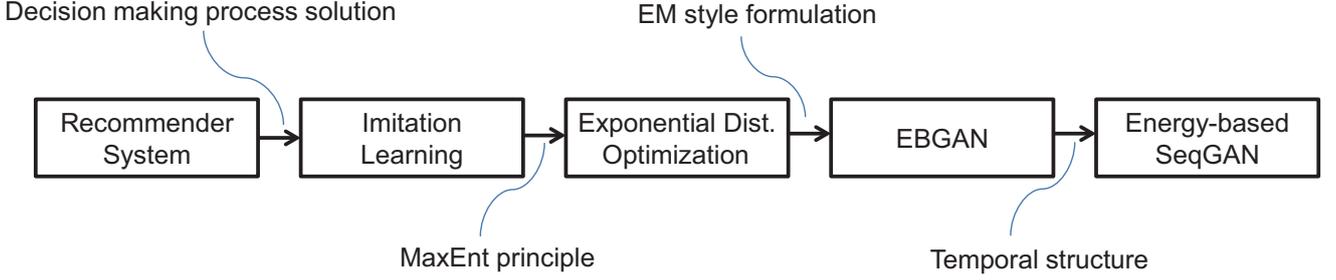

Figure 2: Flowchart of section 4

These steps are very close to expectation maximization algorithm (EM) and approximate inference [3]. Now, imitation learning can be considered as repeating the above process using gradient based optimization.

In (18), $KL(q||P) = \mathbb{E}_{q(S,A)}c(S,A) - H(q) + \log Z$ gives a new objective $\mathbb{E}_{q(S,A)}c(S,A) - H(q)$ for $q$ because $\log Z$ is constant with respect to $q$. In addition, as mentioned above, recommender system can be assumed to have a deterministic transition $a_t = s_{t+1}$, then $A$ can be omitted. Finally, imitation learning on recommender system is formulated as follows.

$$\text{repeat}$$
$$\min_q \mathbb{E}_{S \sim q(S)} c(S) - H(q) \text{ for fixed } c \quad (20)$$
$$\max_c -\mathbb{E}_{S \sim \text{demo}} c(S) + \mathbb{E}_{S \sim q(S)} c(S) \text{ for fixed } q \quad (21)$$

One may observe the similarities between (20), (21) and (3), (4). $c, q$ in (20), (21) correspond to $D$ and $p_G$ in (3),(4) respectively. Substituting negative entropy for regularizer term in (4) makes the objective for $q$ and $p_G$ equivalent, and the objective for $c$ and $D$ are the same except for the minor difference, the margin function $[\cdot]^+ = \max(\cdot, 0)$.

To exploit the learned imitation learning system for the recommendation, the learned distribution of the sequence $P(S) = \exp(-c(S))/Z$ may be used. However, it is intractable to get $P(S)$ when state space is large. Instead, $q(S)$ can be used to recommend items because it becomes close to $P(S)$ provided that the learning process is successful. It matches to using $p_G$ for recommendation.

Applying imitation learning to music recommendation system is reasonable in that constructing the recommender system can be regarded as solving decision making processes. We have showed that our method can be considered as imitation learning by approximating the gradient with approximate distribution. Also it can be interpreted that while imitation learning uses the given feature such as state occurrence, EBGANs determines the energy with implicit feature. i.e EBGANs determines $c^T f$ in (8) without definite feature $f$. Finally, this theoretical connection between EBGANs and MaxEnt imitation learning supports the application of EB-SeqGANs to recommender systems.

## 5 RELATED WORK

The idea of exploiting deep learning and generative modeling for recommender systems has been widely explored. Salakhutdinov et al. [27] developed a recommendation system that can predict movie ratings by using the restricted Boltzmann machine (RBM). Ouyang et al. [23] used autoencoders to predict movie ratings. Van den Oord et al. [32] combined convolutional neural networks (CNNs) and a matrix factorization technique to learn the representation of music data for recommendation. Sahoo et al. [26] utilized the hidden Markov model (HMM) while Hidasi et al. [11] used recurrent neural networks (RNNs) for generative modeling.

There exists prior work that attempted to reveal connections between GANs and other types of learning techniques closely related to imitation learning, although none of these previous attempts were in the context of recommender system design. Ho and Ermon [13] studied the relationship between inverse reinforcement learning (IRL) and the original GANs mentioned in Section 2. Their idea was based on that IRL works by giving rewards to expert policies while penalizing fake policies, which resembles the main idea of GANs. Finn et al. [6] connected the original GANs and the guided cost learning [7], which is to solve the imitation learning problem by using importance sampling and the guided policy search [19]. The key difference between their approach and ours is as follows: Their approach directly approximates the training objective in imitation learning, whereas our method first approximates the gradient of the objective and then approximates the objective using the approximated gradient, as detailed in Section 4. The direct approximation of the objective often tends to cause numerical issues such as overflows and high variances. This difference also allows us to link imitation learning to energy-based GANs rather than the original GANs.

Our idea of replacing the sampling distribution in the imitation learning by an optimization variable was inspired by [10] and has some connection to the actor-critic reinforcement learning [30]. In this paper, we optimize the sampling distribution with sequence generative adversarial network, whereas in [10], the authors used the Stein-variational gradient descent (SVGD) [20]. Applying SVGD to categorical data in recommender systems is difficult, as the kernel functions and the notion of distance in categorical space are often not clearly defined.

## 6 DISCUSSION

As the need for generating sequences of recommendation items emerges in recent recommender systems, adopting GANs (one of the most popular deep generative models) may be a logical step.

In this paper, we have described how to adopt EBGANs to construct such a recommender system and also showed how EBGANs and imitation learning can be related. Applying EBGANs to a recommender system can be interpreted as an instance of applying maximum-entropy imitation learning to the recommender system.

In this paper, we used an RNNs for recommender systems in a GANs framework. Another option to use an RNNs in the context of recommender system design would be to use an RNNs in a supervised learning setting [11]. This can be regarded as a behavior cloning that learns the action at each timestep separately [14]. However, it is known that the behavior cloning suffers from *cascading errors* [14], which can cause a recommender system to return an unlikely sequence of items as the error accumulates. In the case of a music recommender system, avoiding such a problem is particularly critical because many users often turn on the so-called auto-play functionality. Our proposed method can be regarded as imitation learning and can learn an entire trajectory of recommendations, rather than individual ones. We thus expect that our approach will suffer less from the cascading error issue.

Our next step will be testing the proposed method with various recommendation data for thorough performance evaluation. As mentioned previously, our methodology is not limited to music recommender systems but is applicable to other types of sequential recommender systems (e.g., click stream analysis and recommendation). Of note is that the popular top@k recall metric [11] will not be ideal for assessing the performance of sequential recommender systems. This metric is not related to cascading errors as the previous history is not generated by the recommender system itself but always given by data. Consequently, we will need an alternative metric to better assess the performance of a recommender system that returns a sequence of recommendations.

We may encounter other challenges when testing the proposed approach with real data. Examples include the strong dependence of the sequence GAN components on pretraining and slow convergence to equilibria [25]. We expect that employing appropriate techniques such as trust region policy optimization (TRPO) will be helpful for addressing such challenges [28] [13].